\documentclass{iopart}
\usepackage{graphicx}
\usepackage{amsfonts}
\usepackage{color}
\usepackage{braket}

\usepackage{epstopdf}
\usepackage{cite}
\usepackage{multicol}
\newcommand{\overbar}[1]{\mkern 1.5mu\overline{\mkern-1.5mu#1\mkern-1.5mu}\mkern 1.5mu}

\def\gsim{\mathrel{\rlap{\lower4pt\hbox{\hskip1pt$\sim$}}\raise1pt\hbox{$>$}}}

\begin{document}

\title{Rewiring Stabilizer Codes}

\author{Kristina R. Colladay and Erich J. Mueller}
\address{Laboratory for Atomic and Solid State Physics, Cornell
University, Ithaca NY 14853}
\eads{\mailto{krc76@cornell.edu},\mailto{em256@cornell.edu}}
\date{\today}

\begin{abstract}
We present an algorithm for manipulating quantum information via a sequence of projective measurements.  We frame this manipulation in the language of stabilizer codes: a quantum computation approach in which errors are prevented and corrected in part by repeatedly measuring redundant degrees of freedom.  We show how to construct a set of projective measurements which will map between two arbitrary stabilizer codes.  We show that this process preserves all quantum information.  It can be used to implement Clifford gates, braid extrinsic defects, or move between codes in which different operations are natural.  
\end{abstract}

%
\vspace{2pc}
\noindent{\it Keywords}: Quantum Computing, Stabilizer Codes, Code Conversion
%
%
%
%

\submitto{\NJP}

\maketitle

\section{Introduction}
Although there is broad agreement that quantum mechanics can provide an important resource for computation, 
the community continues to search for the best way to exploit that resource.  Here we describe a tool
which can be incorporated into a number of quantum information processing architectures, and we show how it can be applied to solve important problems such as giving access to a universal set of transversal gates and braiding non-abelian anyons.

Our approach is framed in the language of quantum stabilizer codes.  These are approaches to quantum error correction in which 
$k$ logical qubits are stored in $n$ physical qubits (quantum spins or other two level systems): the remaining degrees of freedom are restricted by requiring that the allowed wavefunctions (those in the codespace) are 
eigenstates of $n-k$ given stabilizer operators.  These stabilizer operators define the code.  

Given two arbitrary stablilizer codes, we show how to construct a sequence of 
measurements which map the codespace of one code into the codespace of the other while preserving the quantum information.  Under appropriate circumstances this mapping is fault-tolerant.  The utility of such rewiring has been recognized by the community, and our arguments build on ideas of ``code deformation" \cite{Bombin2009, Bombin2011} (which involves small rewirings of topological codes) and ``code conversion"  \cite{Stephens2008, Hill2011}.  This latter strategy has also been termed ``code switching"
\cite{Nikahd2017, Yoder2016}, and typically is based on a sequence of unitary gates. Our contribution is the construction of a general algorithm for finding a sequence of projective measurements which allow arbitrarily large deformations.  As with the previously studied cases, rewiring can be used to implement quantum gates:  a cyclic rewiring generically acts as a rotation in the codespace.   We explicitly construct the unitary matrix that corresponds to the gate.

One motivator for mapping between codes is the realization that
the amount of resources needed to perform different gates depends on the code.  Recently, Paetznick and Reichardt  \cite{Paetznick2013} noted that this disparity can be taken advantage of by mapping between codes.  Specifically, a mapping can thereby produce a universal set of gates which have a particularly simple structure, described as ``transversal".  Transversal operations are naturally fault-tolerant, but there is no single code that admits a universal set of transversal gates \cite{Eastin2009}.  To overcome this, Anderson, et. al. \cite{Anderson2014} proposed a quantum circuit which maps between the 7-qubit Steane code and the 15-qubit Reed-Muller code. The former admits transversal Cifford gates, while the latter admits transversal T and control-control-Z gates.  Subsequently, several authors proposed other circuits for this mapping \cite{Hwang2015,Quan2017}.  Our algorithm provides a systematic approach to this, and related, problems.  We illustrate its utility by producing a mapping between the Steane and Reed-Muller codes.  The resulting circuit is particularly simple, involving only measuring stabilizer generators of the other code, and is fault-tolerant.

Our algorithm fits in with a long tradition of using measurement to manipulate quantum information.
 Measurement is a key piece of any quantum circuit, and there are well known computing architectures which consist primarily of repeated measurements \cite{Gottesman1999,Raussendorf2001,Knill2001,Nielsen2003,Leung2004,Verstraete2004,Gross2007,Gross2007a,Briegel2009,Aliferis2004}.  These algorithms are often discussed within the stabilizer formalism \cite{Aliferis2004} and can be interpreted as rewiring stabilizer codes.  In this context, we emphasize that our innovation is not the basic idea that one can use projective measurements to map between codes, but rather the explicit construction of the sequence of measurements.

Beyond its application to quantum computing, our algorithm can be used to enable the study of novel collective effects of interest to condensed matter physics.  A physical realization of a quantum stabilizer code can be considered a Hamiltonian system where the Hamiltonian is simply the projector onto the codespace.  Kitaev has argued that this mapping allows the observation of anyons (excitations which behave as particles with statistics that are neither fermionic nor bosonic \cite{Kitaev2003}).  Using our code rewiring algorithm, we show how to braid non-abelian twist defects, allowing the first direct observation of non-abelian quasiparticle statistics.  Such ``quantum emulation" experiments would be highly impactful.

\section{Algorithm}\label{algo}
\subsection{Stabilizer Codes}

A stabilizer code stores $k$ logical qubits in $n$ physical qubits by specifying that a physical state is an eigenstate of $n-k$ given stabilizer operators (generators) with eigenvalue $1$.   The space of physical states is denoted the codespace.  The stabilizer generators ($g_0,g_1,\ldots g_{n-k-1}\in G$) are typically taken to belong to the Pauli group -- meaning that they are products of Pauli operators ($I,X,Y,Z$) acting on the physical qubits.  They therefore all have eigenvalues $\pm 1$.  The generators should be independent, and the space generated by their products is denoted $S$.  For the codespace to be non-trivial, the stabilizers (and hence the generators) must commute with one-another.  The generators for a stabilizer code are not unique:  the same set $S$ is generated if $g_i g_j$ replaces $g_i$, for any $j$.

In addition to the stabilizers, one can define logical operators, which map the codespace onto itself.  The space of logical operators is generated by $2k$ members of the Pauli group, which can be labeled 
$\bar{X}_1,\cdots \bar X_k,\bar Z_1,\cdots \bar Z_k$, 
with $\bar X_j$ anticommuting with $\bar Z_j$, but commuting with all other logical generators.  The labeling of the logical operators is not unique.  Any member of the Pauli group that commutes with the stabilizers, but is not itself a stabilizer, is a logical operator.

\subsection{Moving in the space of stabilizer codes}\label{onestep}
Consider two stabilizer codes  $S,S^\prime$ that differ by only one anti-commuting generator: $S$ is generated by $G=\{g_0,g_1,\cdots g_{n-k-1}\}$, while $S^\prime$ is generated by $G^\prime=\{g_0^\prime,g_1,g_2,\cdots g_{n-k-1}\}$, with anticommutator $\{g_0,g_0^\prime\}=0$.    As is readily verified by its action on the generators, the unitary operator $U=(1+g_0^\prime g_0)/\sqrt{2}$ maps the codespace of $S$ to the codespace of $S^\prime$.  One can further see that when acting on a state $|\psi\rangle$ in the codespace of $S$,
\begin{equation}\label{unitary}
 U |\psi\rangle = \sqrt{2} P_1 |\psi\rangle =\sqrt{2}  g_0   P_{-1} |\psi\rangle,
\end{equation}
where  $P_{\pm 1}=(1\pm g_0^\prime)/2$ are projectors into the space spanned by the $\pm 1$ eigenstates of $g_0^\prime$.  This relationship suggests an algorithm.  Starting from a state $|\psi\rangle$, one measures $g_0^\prime$.  If the result of the measurement is $-1$, one applies $g_0$ to the new state, otherwise one leaves it alone.  Due to Eq.~(\ref{unitary}), this procedure is completely equivalent to the unitary transform, but is often simpler to implement.

One can chain these operations together -- one-by-one changing the stabilizer generators.  In the next section we show how to construct a sequence of measurements that map between any two given stabilizer codes.  If the final code is the same as the initial code, then we thereby apply a gate (given by the product of the $U$'s in Eq.~(\ref{unitary})).  This feature is used in a number of algorithms, such as topological \cite{Bombin2009, Bombin2011}, and teleportation\cite{Gottesman1999} based computing schemes.

The operator $U$ in Eq.~(\ref{unitary}) maps the Pauli group onto itself, and is therefore described as a Clifford operator.  Clifford operators are insufficient for universal quantum computation, and a quantum circuit only involving Clifford operators can be efficiently simulated on a classical computer \cite{Nielsen2010}.  Relaxing the constraint that the generators belong to the Pauli group opens up the ability to generate arbitrary gates.  

Although the arguments have largely appeared elsewhere, in  \ref{proofs} we give the proofs that $U$ has all of these properties.  Note, if one has a non-measurement way to apply $U$, such gates could also be used as part of the rewiring.  

\subsection{Constructing a sequence of stabilizer codes}\label{path}
Given two stabilizer codes $S,S^\prime$, we wish to construct a sequence of stabilizer codes $S_0,S_1,\cdots,S_N$, such that $S_0=S$, $S_N=S^\prime$, and $S_j$ differs from $S_{j-1}$ by only one anticommuting generator.  By the arguments in Sec.~\ref{onestep} one then can map between $S$ and $S^\prime$ by performing $N$ measurements.

To facilitate constructing this sequence of stabilizer codes, we first make use of the fact that the generators are not unique, and find a set of generators of $S$ and $S^\prime$ such that each of the generators fall into one of three blocks, denoted $G_A,G_B,G_C$ and $G^\prime_A,G^\prime_B,G^\prime_C$.  The first blocks are identical: $G_A=G_A^\prime$.  The second blocks, $G_B$ and $G_B^\prime$ contain the same number of elements, and have the property that the generators in $G_B$ commute with members of $S^\prime$, but are not in $S^\prime$.  Conversely, those in $G_B^\prime$ commute with the members of $S$, but are not in $S$.  Thus the elements of $G_B$ are logical operators for the code defined by $S^\prime$, and vice versa.  The elements in the third block are in one-to-one correspondence, and for each $g_j\in G_C$, there exists a single element $g_j^\prime \in G_C^\prime$, such that $g_j$ and $g_j^\prime$ anticommute.  The element $g_j$ however, commutes with all other elements of $G^\prime$ (and likewise for $g_j^\prime$).  We claim that one can always find generators satisfying these conditions, and in Sec.~\ref{construct} we provide a constructive algorithm for finding these generators.

For each element $g_j\in G_B$ our construction also gives a complementary operator $g_j^{(c)}$ with the property that $g_j^{(c)}$ anticommutes with $g_j$, but commutes with all other elements of $G$.  Furthermore $g_j^{(c)}$ commutes will all elements of $G^\prime$.  Similarly, for each $g_j^\prime\in G_B^\prime$ we find a complementary operator $g_j^{\prime(c)}$ with analogous properties.

Given this construction, we can then generate the sequence of stabilizer codes $S_0,S_1,S_2,\cdots S_N$. If the number of generators in blocks $G_A,G_B,G_C$ are $a,b,c$, then we require $N=2 b+c$ steps.  It will require one step to replace an element of $G_C$ with one of $G_C^\prime$, and two steps to replace an element of $G_B$ with one of $G_B^\prime$.  The required operation for replacing elements of $G_C$ is simple: one just measures the elements of $G_C^\prime$. 
In order to replace $g_j\in G_B$ with $g_j^\prime \in G_B^\prime$, one first measures the product $g_j^{(c)} g_j^{\prime(c)}$, then measures $g_j^\prime$.  Each of these steps is of the form detailed in Sec.~\ref{onestep}.

Note if $G_B$ is empty, then the two codes $S$ and $S'$ are different gauge fixings of a single subsystem code.  In that case, our algorithm reduces to a standard gauge fixing approach.  See Sec.~\ref{sub}.

In \ref{trivial} we give some simple examples to illustrate the mechanics of this procedure. 
Section~\ref{examples} further explores the utility of  these mappings.

\subsection{Construction of the generators and complementary operators}\label{construct}
In this section we explicitly construct generators of the form described in Sec.~\ref{path}.
Given an arbitrary set of generators, $G$ and $G'$ of $S$ and $S^\prime$, we first construct a connectivity matrix $M$, whose $(i,j)$'th element is 1 if the $i$'th member of $G$ anticommutes with the $j$'th member of $G^\prime$.  Otherwise that element is zero.  For codes with $n$ physical qubits and $k$ logical qubits, $M$ will be a $(n-k)\times (n-k)$ matrix.  As already emphasized, the generators are not unique, and the same set of stabilizers are formed if one replaces any one generator by its product with another.  Such a replacement in $G$ corresponds to adding the $j$-th row of $M$ to the $i$-th row mod(2).  A similar replacement in $G^\prime$ corresponds to adding columns mod(2).  By using the same techniques used for row reduction, one can thereby find a set of generators for which $M$ is diagonal (with zeros and ones along the diagonal).  The generators $g_i$ and $g_i^\prime$ with $M_{ii}=1$ correspond to block $C$, as defined in Sec.~\ref{onestep}.  

Consider any $g_j \notin G_C$.  Because of the structure of $M$, it commutes with all of the generators of the stabilizer group of $S^\prime$.  Thus it is either a member of $S^\prime$ or logical operator of $S^\prime$.  If it belongs to $S^\prime$, then it equals a product of the members of $G^\prime$ -- and one can always construct a set of generators for $S^\prime$ such that $g_j\in G^\prime$, giving us blocks $G_A$ and $G_A'$.  This rearrangement of $G^\prime$ does not require replacing any of the elements of block $C$.  Finally, all of the remaining generators are logical operators for the other code, and hence are in blocks $G_B$ and $G_B'$.

We further transform the generators, based upon the operators in blocks $G_B$ and $G_B'$.  Each $g_j\in G_B$ is a logical operator for $S^\prime$, and hence there is a complementary logical operator $g_j^{(c)}$, that anticommutes with $g_j$. 
By construction, $g_j^{(c)}$ commutes with all elements of $G^\prime$ (and hence the elements of $G_A$).  The complementary operators can always be chosen to commute with each other:
  If $\{g_j^{(c)},g_k^{(c)}\}=0$ for some $g_k \in G_B$, one simply takes $ g_j^{(c)} \to g_j^{(c)} g_k$.   After this manipulation, one can further transform the generators so that $g_j^{(c)}$ commutes with all if $g_k\in G_B$ or $G_C$, for $k\neq j$:   If $g_k\in G_C$ anticommutes with $g_j^{(c)}$, we take $g_k\to g_k g_j$.  This transformation does not change any of the other commutation relations.
Note, as illustrated in Sec. \ref{examples}, one can map between codes with different numbers of qubits by simply appending auxillary qubits to the shorter code with trivial local stabilizers.

\subsection{Distance Bounds}\label{mindist}
In commonly-studied noise models, noise acts independently on physical qubits.  Consequently, one measure of the robustness of the  code is its 
{\em distance} -- the lowest weight operator that performs a logical X- or Z- operation.  Here {\em weight} is the number of qubits which are acted on by the operator.  A code of distance $d$ can correct errors of weight $t$ or less, where $d=2t+1$.

Our basic algorithm can produce  intermediate codes whose distance is smaller than that of $S$ and $S^\prime$.  ~\ref{fail} gives an explicit example.  In Secs.~\ref{sub} and \ref{sub2} we  quantify this issue by deriving lower bounds on the distances of the intervening codes generated by our algorithm when mapping between $S$ and $S^\prime$.  As recently shown by Huang et al. \cite{Huang2018}, these lower bounds are worst case scenarios, and after adding appropriate ancilla qubits, there always exists a path in which the intermediate codes have distance no smaller than $S$ or $S^\prime$.  Moreover, Huang et al. find the remarkable result that a random path yields high distance codes with high probability (and this probability can be made arbitrarily large by adding more ancillas).

\subsubsection{Subsystem Codes.}\label{sub}

We first consider the case where $G_B$ is empty, and
one only needs to measure stabilizers belonging to the code to which we wish to map.  In this case our algorithm may be  interpreted as a gauge fixing procedure on a subsystem code \cite{Poulin2005, Kubica2015, Bombin2010a,Bombin2016}.  Below we describe subsystem codes and outline this procedure.  In particular we show that if $G_B$ is empty, then all codes in the path between $S$ and $S'$ are gauge fixings of a single subsystem code.  One consequence is that the distances of each of these codes are bounded below by the distance of that subsystem code. 

A subsystem code stores $k$ logical qubits in $n$ physical qubits.  States in the codespace are eigenstates of $s$ independent stabilizer operators (forming a group $S$ and having generators $G$), where $s<n-k$.  This leaves $r=n-k-s$ degrees of freedom which are not used to encode any information.  These $r$ degrees of freedom are called gauge qubits, and they can be freely manipulated.  The gauge group ${T}$ is composed of Pauli operators that commute with both the logical operators and the stabilizers of the subsystem code.  
This is the set of operators which cannot
disturb the encoded information.  
The gauge group of a subsystem code is generated by $s+2 r$ elements -- the generators of the stabilizer group and the generators of the ``logical operators" that act on the gauge qubits.
Gauge fixing amounts to creating a stabilizer code whose generators consist of $r$ independent commuting elements of ${T}/{S}$ along with the elements of ${G}$.  Choosing those $r$ elements in different ways can produce distinct stabilizer codes.  
Given a state stored in a subsystem code, its gauge-fixed variant is produced by simply measuring the relevant gauge operators.

Consider two stabilizer codes with stabilizer generators $G$ and $G^\prime$, decomposed into blocks $A$, $B$, and $C$, as in Sec.~\ref{path}.  Recall that the elements of the two $A$ blocks are identical: $G_A=G_A^\prime$, and the elements of the two $C$ blocks are in one-to-one correspondence with $g_j\in G_C$ anticommuting with one $g_j^\prime \in G_C^\prime$ and commuting with all other elements of $G_C^\prime$.  If the $B$ block is empty, we can construct a subsystem code by taking its stabilizer group to be generated by $G_A=G_A^\prime$.  We use $G_C$ and $G_C^\prime$ as the generators of the gauge group (along with $G_A$).   The two stabilizer codes are created from this subsystem code via gauge fixing.  Our algorithm then reduces to the standard gauge-fixing process, and each code along the path represents a different gauge fixing.  A consequence is that if the subsystem code has distance $d$, then we are guaranteed that each code visited during the conversion will have at least distance $d$.  This distance, for example, can be calculated using the approach in \cite{Knapp2018}.

When $G_B$ and $G_B^\prime$ contain $b>0$ elements, the most naive generalization of this procedure fails.  Namely, consider a subsystem code with stabilizer group $G_A$ and gauge group containing the union of $G_B,G_B^\prime, G_C,$ and $G_C^\prime$.  If $b$ is the number of elements in $G_B$ and $c$ is the number of elements in $G_C$, this gauge group contains at least $2b+c$ linearly independent commuting elements -- namely the set $G_B\cup G_B^\prime\cup G_C^\prime$.  Thus it encodes at most $n-(a+2b+c)$ logical qbits, which is smaller than the $n-(a+b+c)$ qbits encoded by the two original stabilizer codes.

\subsubsection{Generic Case.}\label{sub2}

Here we generalize the argument of Sec.~\ref{sub} to produce a general bound on the distances of our codes.  In particular, if the B-blocks of $S$ and $S^\prime$ contain $b$ elements each, then we construct $2^b$ subsystem codes.  We argue that any stabilizer code generated by our procedure must be a gauge fixing of one of these.  In the special case where $b=0$, this procedure generates the subsystem code in Sec.~\ref{sub}.  The distance of any code generated by our procedure is then bounded below by the smaller of the distances of all these subsystem codes.  For modest $b$ the task of enumerating these subsystem codes and finding their distances is reasonable.

For this construction we take the B-blocks to contain $G_B = \{Z_1', ..., Z_b' \}$ and $G_B' = \{Z_1, ..., Z_b\}$ where $Z_j'$ and $Z_j$ are logical operators for $S$ and $S'$.  Note, this notation does not imply that these necessarily correspond to logical $Z$.  The complementary logical operators are $\{X_1', ..., X_b'\}$ and $\{X_1, ..., X_b\}$.   We further define the set $\overbar{G_B} = \{X_1X_1', X_2X_2', ..., X_bX_b'\}$ and  the $2^b$ sets $G_B^s$ made by choosing $h_s$ elements from $G_B$ and $b-h_s$ from $G_B^\prime$.  Here $s$ labels which of these choices are made.  We then construct the subsystem code with stabilizer group $G_A$, and gauge group generated by $G_A, G_B^s, \overbar{G_B}, G_C, G_C^\prime$.  
Any code traversed in our procedure will have its stabilizer generators in the gauge group of one of these codes, and hence is a gauge fixing of that code.

\subsection{Fault Tolerance} \label{ft}

Intuitively, a measurement of the form of those in Sec.~\ref{onestep}, is  fault-tolerant if errors that occur during the measurement do not degrade the error-protective properties of the code.  For a code of distance 3 or higher (ie. one in which at least a single qubit error can be corrected), this intuitive requirement can be satisfied by the property that no
single error during the measurement can propagate to more than one qubit on the data\cite{Nielsen2010}. The cat state method, as introduced by Shor \cite{shor} and nicely discussed by Aliferis, Gottesman, and Preskill \cite{Aliferis2006}, provides one  approach to meeting this requirement.  The measurement is described in detail in \ref{ftmeasurement}.

\subsection{Implementing Constraints}
In practical applications, the physical apparatus may introduce constraints on the operators measured.   For example, one may only be able to carry out single qubit measurements, or measurements on sets of qubits that are connected by hard-coded wires.  In the presence of such constraints, it is no longer possible to map between any two arbitrary stabilizer codes.  The mapping may also be asymmetric: the constraints may be satisfied in mapping from A to B, but not in mapping from B to A.  This latter setting is the domain of ``one-way" quantum computing \cite{Gottesman1999}.

One defines the constraints through the set  $W$ of all measurable operators.  Given stabilizers codes $S$ and $S^\prime$ generated by $G$ and $G^\prime$, a necessary, but not sufficient, condition for the existence of a rewiring path is that $G^\prime$ is a subset of the group generated by $W\cup G$.  For small systems, constrained paths can be found via an exhaustive search.

Note in many cases, measuring high-weight stabilizers is problematic.  If one constrains the weights of the generators to be measured along the path from one code to another, such a path is not guaranteed to be generated by the algorithm.

\section{Applications}\label{examples}

In this section we illustrate the utility of our algorithm by using it in physically relevant cases: code conversion for universal transversal computing and braiding defects and twists in topological codes.

\begin{table}
\caption{\label{sr}Generators for Steane code ($g_0 - g_5$) and Reed-Muller codes ($g_0' - g_{13}'$).  Additional qubits with arbitrarily chosen stabilizer generators  ($g_6 - g_{13}$) are appended to the Steane code for the conversion.}
\begin{indented}
\item[]\begin{tabular} {@{}lll}
\br
&Steane
&
Reed-Muller\\
$j$&$g_j$&$g_j^\prime$\\
\mr
0& $X_1X_3X_5X_7$ & $X_1X_3X_5X_7X_9X_{11}X_{13}X_{15}$ \\
1& $X_2X_3X_6X_7$ & $X_2X_3X_6X_7X_{10}X_{11}X_{14}X_{15}$\\
2&$X_4X_5X_6X_7$& $X_4X_5X_6X_7X_{12}X_{13}X_{14}X_{15}$\\
3&$Z_1Z_3Z_5Z_7$ & $X_8X_9X_{10}X_{11}X_{12}X_{13}X_{14}X_{15}$\\
4&$Z_2Z_3Z_6Z_7$ & $Z_1Z_3Z_5Z_7Z_9Z_{11}Z_{13}Z_{15}$ \\
5&$Z_4Z_5Z_6Z_7$ & $Z_2Z_3Z_6Z_7Z_{10}Z_{11}Z_{14}Z_{15}$\\
6&$Z_8$& $Z_4Z_5Z_6Z_7Z_{12}Z_{13}Z_{14}Z_{15}$\\
7&$Z_9$ & $Z_8Z_9Z_{10}Z_{11}Z_{12}Z_{13}Z_{14}Z_{15}$\\
8&$Z_{10}$ & $Z_1Z_3Z_9Z_{11}$\\
9&$Z_{11}$ & $Z_2Z_3Z_{10}Z_{11}$\\
10&$Z_{12}$ & $Z_3Z_7Z_{11}Z_{15}$\\
11&$Z_{13}$ & $Z_1Z_3Z_5Z_7$\\
12&$Z_{14}$ & $Z_2Z_3Z_6Z_7$\\
13&$Z_{15}$ & $Z_4Z_5Z_6Z_7$\\
\br
\end{tabular}
\end{indented}
\end{table}

\subsection{Universal transversal computing}\label{utc}
The [[7,1,1]] Steane code is a 7-qubit code that supports transversal Clifford gates, and the  15-qubit [[15,1,3]] quantum Reed-Muller code supports several transversal non-Clifford gates which complement those of the Steane code.   Thus  \cite{Hwang2015,Quan2017,Anderson2014} suggested sequentially mapping between these codes to produce a universal set of transversal gates.  The generators for these codes are given in Table~\ref{sr}, where we have appended extra bits to the Steane code.

Without any reference to the structure of the codes, our algorithm gives a mapping between them (see \ref{storm} for details).  For example, we find that 7 operators need to be measured to map from the Steane to the Reed-Muller code: $g_0^\prime,g_1^\prime,g_2^\prime,g_3^\prime,g_8^\prime,g_9^\prime,g_{10}^\prime$.  Remarkably, these are all stabilizer generators of the Reed-Muller code, and hence no additional hardware is required, beyond what one already needs for implementing error correction. These operators can be measured in any order desired, or even simultaneously.  Similarly one can map back to the Steane code by only measuring stabilizers:  $g_0,g_1,g_2,g_6,g_8,g_9,g_{10}$.  The resulting round-trip acts as the identity operator.  The distance of the code never falls below 3 so, as discussed in Sec.~\ref{ft}, the process is fault-tolerant.

Due to its importance, the literature contains several other approaches to map between the Steane and Reed-Muller codes.  Hwang et al. produced a circuit based on the particular structure of these codes and the fact that they can both be transformed into a common canonical form \cite{Hwang2015}.  Anderson, Duclos-Cianci, and Poulin made the insightful observation that these two codes could be considered to be identical subsystem codes -- with additional stabilizers added to fix the gauge  \cite{Anderson2014}.  They were able to use this structure to generate a map between the codes.  Bombin also made similar observations \cite{Bombin2015}.  Quan et al. later extended this idea \cite{Quan2017}.  Our approach is in the same category as these gauge-fixing methods, but has the added benefit that it automatically finds the minimal set of operators which need to be measured.  It is purely algorithmic and does not require any insight.

As a related example, Huang and Newman used our algorithm to create a mapping between the 5-qubit code and the Steane code \cite{Huang2018}.

\subsection{Moving defects in topological codes}\label{top}

Topological codes are stabilizer codes where the spins are arranged on a lattice,  the stabilizer generators are local (meaning they only involve spins which are near one-another), but the logical operators are non-local.   These codes are particularly robust against local noise sources.  They are also of great intellectual interest, as they have connections to gauge theories, spin liquids, and topological order.  These codes are typically translationally invariant, but it can be advantageous to introduce ``extrinsic defects," which locally disrupt the wiring.  One  can implement gates by deforming the code so that these defects move around one-another \cite{Fowler2012}.  Here we apply our algorithm to the problem of moving these defects -- reproducing known protocols for moving ``$e$" and ``$m$" type defects in surface codes, and finding new protocols for converting between these defects, and moving ``twist defects".  The latter is valuable beyond quantum information processing, 
as moving twist defects around one-another is equivalent to braiding Majorana fermions, which would allow the direct observation of excitations with non-Abelian statistics.

\begin{figure}[t]
\begin{indented}
\item[]\includegraphics[width = 0.8\columnwidth]{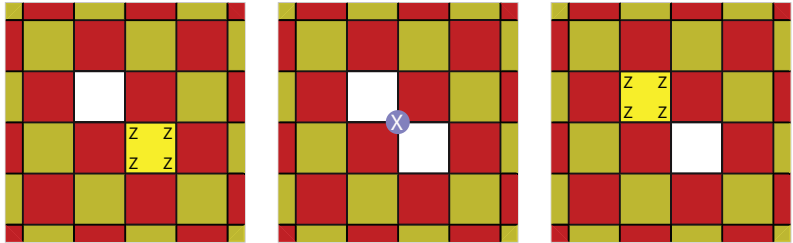}
\end{indented}
\caption{Transporting an $e$ type defect in the surface code.  Qubits sit at the vertices of the square lattice, yellow (light) and red (dark) plaquettes denote stabilizers corresponding to the product of $Z$ or $X$ operators on the four qubits at the corners. The white squares represent the absence of a stabilizer.  To convert from the code on the left to the one on the right, one first measures the $X$ operator highlighted in the middle panel, then the highlighted stabilizer in the right panel.}
\label{fig: zzType}
\end{figure}

We will explicitly consider the toric/surface code.  There are several conventions for defining this code.  We follow \cite{Bombin2010}, and place the qubits on the vertices of a checkerboard lattice with alternate plaquettes colored yellow and red (Fig. \ref{fig: zzType}).  For each yellow plaquette there is a stabilizer corresponding to the product of $Z$'s on the four qubits at the vertices.  For each red plaquette there is another stabilizer corresponding to the product of $X$'s.  The number of logical qubits stored by the code depends on the boundary conditions and the topology of the surface tiled by the qubits.

The simplest defect involves ``removing" one of the stabilizers -- meaning that one removes that generator from G, reducing the constraints on the codespace.  In the presence of appropriate boundaries, this extrinsic defect yields one additional logical qubit.  This defect is referred to as an $Z$ or $X$ defect (or equivalently $e$ and $m$) depending if the removed stabilizer is a product of $Z$'s or $X$'s.  Bombin \cite{Bombin2009,Bombin2011}  uses the phrase ``code deformation" to describes the operations required to move these defects, and the algorithm is explained in detail by Fowler \cite{Fowler2012}.  Our general code rewiring algorithm can be used to find these operations.

Consider for instance the setup in Fig.~\ref{fig: zzType} where one wishes to move a $Z$ defect (shown as a white square in the leftmost panel) to a diagonally adjascent site (rightmost panel).  The two codes differ by only one generator: $g_0=Z_1Z_2Z_3Z_4$, $g_0^\prime=Z_4 Z_5 Z_6 Z_7$, where $Z_4$ acts on the shared qubit.  These are logical operators for the other code (ie. lie in the B blocks).  The complementary logical operators involves a product of $X$ operators extending from the each of the defects to the boundary.  The product of these logical operators is simply $X_4$.  Thus, following the prescription in Sec.~\ref{onestep}, one rewires the code by first measuring $X_4$ then measuring $g_0^\prime$.  This protocol coincides with the one in \cite{Fowler2012}.

\begin{figure}[t]
\begin{indented}
\item[]\includegraphics[width = 0.8\columnwidth]{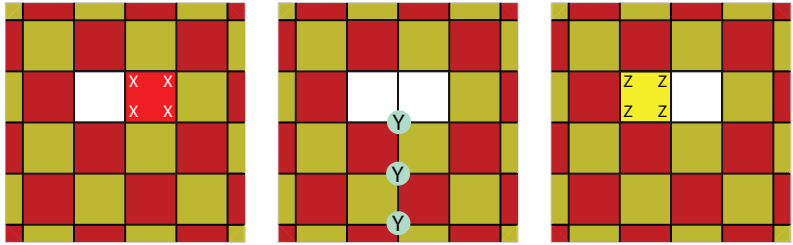}
\end{indented}
\caption{Changing an $e$ type defect into an $m$ type defect. This conversion requires measuring a non-local operator corresponding to a product of Y operators extending from the defects to the edge of the sample. }
\label{fig: zxType}
\end{figure}

As a second example, consider Fig.~\ref{fig: zxType} where one wishes to convert a $Z$ defect into an $X$ defect on a neighboring plaquette.  Again the codes differ in only one generator, and the generators are logical operators for the other code.  The complementary operators this time are a string of $X's$ and a string of $Z$'s extending to a boundary.  Their product is a string of $Y$'s.  Thus rewiring this code requires measuring a non-local quantity.  In many topological quantum computer architectures this is impractical, as one wishes to avoid measuring non-local operators.  In such a case, one could first move the defect near a boundary, where the string becomes short, at the cost of reducing the distance of the code.

Finally we consider a ``twist" defect, illustrated in Fig. 3.  Two of the square plaquettes are replaced by pentagons, and the plaquettes between them become rhombuses.  As illustrated in the figure, the stabilizer associated with a pentagon involves measuring a product of three $X$ operators, a $Y$ and a $Z$ or the product of three $Z$ operators, a $Y$ and a $X$.  The stabilizer associated with a rhombus involves the product of two $Z$'s and two $X$'s.  Twist defects are important for several reasons.  For example, an $X$ defect can be converted into a $Z$ defect by moving it through the rhombus cells -- a procedure that only requires local measurements.  Given appropriate boundary conditions, adding a twist defect yields one additional logical qubit.

We wish to consider how to deform the code in order to move one of the pentagons.  For example, in Fig. 3 we illustrate a move in which the defect is made shorter.  In this case, the code changes by two generators: On the left,
$g_0 = Z_1Z_2X_4X_5$ and $g_1 = X_2X_3Z_5Y_6X_7$.  On the right, $g_0' = Z_1Z_2X_4Y_5Z_6$ and $g_1' = X_2X_3X_6X_7$.  Following the algorithm in Sec.~\ref{algo},
we find a new set of generators $G=\{g_0,g_1 g_0\},G^\prime=\{g_0^\prime, g_1^\prime g_0^\prime\}$, where $g_0g_1 = g_0'g_1'$, and $g_0$ anticommutes with $g_0^\prime$.  Thus to convert between the codes one need only measure $g_0^\prime$.

\begin{figure}[t]
\begin{indented}
\item[]\includegraphics[width = 0.55\columnwidth]{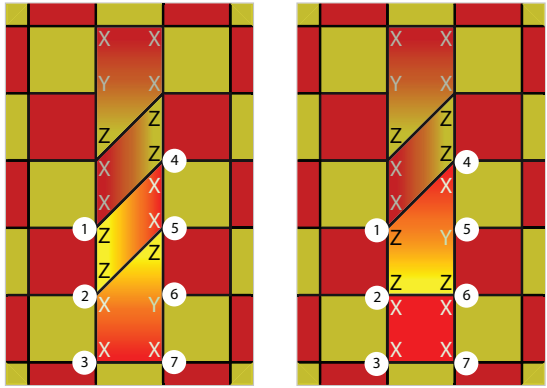}
\end{indented}
\caption{Shortening a twist.  Twist defects, as illustrated in this figure, contain stabilizers corresponding to products of four or five operators.
The  ends of these defects have properties analogous to Majorana fermions.  One converts the code on the left to the one on the right by measuring the operator $X_2 X_3 X_6 X_7$.   Our algorithm similarly gives the procedure for other moves.  }
\label{fig: twist}
\end{figure}

These manipulations allow for operations which can be interpreted as braiding quasiparticles with unusual quantum statistics.  As discussed by Kitaev \cite{Kitaev2003}, the codespace can be identified with the ground-space manifold of a Hamiltonian which is the sum of the stabilizer generators.  Turning off one of the generators, say $Z_0$, enlarges the space.  States that are eigenstates of $Z_0$ with eigenvalue $-1$ are identified with excited states of the Hamiltonian and are said to contain a quasiparticle at that location.  Moving a $Z$ defect around a $X$ defect produces a control-not gate \cite{Fowler2012}, which in this context can be interpreted as a phase that is contingent on the presence of the quasiparticles.  Hence, there is a phase generated by moving one quasiparticle around another, which is the defining property of an anyon.  The twist defects are even more interesting, as in this mapping the pentagons play the role of particles with non-Abelian statistics (described either as Ising anyons or Majorana fermions) \cite{Bombin2010}.

\section{Summary and Outlook}

Quantum codes, first introduced for error correction, become more powerful with operations that map between them.  The way information is encoded influences what operations are most accessible; moreover, gates can be applied through the act of mapping between codes.  This code-conversion paradigm is at the heart of one-way quantum computing and topological quantum computing.  In this paper, we introduce a general-purpose algorithm for mapping between two arbitrary stabilizer codes, often in a fault-tolerant manner.  We illustrate two applications of the algorithm:  mapping between the Steane and Reed-Muller codes and moving defects in topological codes.  

Our algorithm provides a means for creating a mapping between arbitrary stabilizer codes.  Such mappings are not unique, and depending on which path is taken, a different logical gate can be produced.  It would be exciting to extend our approach and gain the ability to choose which logical operation is performed: i.e. given a desired logical operator, how does one construct a sequence of measurements that performs the desired gate?  Similarly, it would be useful to develop approaches that allow the incorporation of constraints, such as locality or code distance
\cite{Huang2018}.   

\section*{Acknowledgements}
We thank Bryan Eastin for critical comments, including providing the example in ~\ref{fail}.
This material is based upon work supported by the National Science Foundation Grant No PHY-1508300, and the ARO-MURI Non-equilibrium Many-body Dynamics grant (W911NF-14-1-0003).

\appendix \label{append}

\section{Properties of $U$}\label{proofs}
Here we establish that $U=(1+g_0^\prime g_0)/\sqrt{2}$ in Eq.~(\ref{unitary}) is a unitary Clifford rotation, and we verify its relationship to the projectors $P_{\pm1}=(1\pm g_0^\prime)/2.$ 
We rely on the fact that $g_0^\prime$ and $g_0$ are Pauli operators that anti-commute, $\{g^\prime_0,g_0\}=0$.
The fact that $U$ is unitary follows from writing
\begin{equation}
UU^\dagger = (1+g^\prime g)(1+g g^\prime)/2=(1+ g^\prime g g g^\prime +\{g^\prime,g\})/2,
\end{equation}
where we have neglected the subscript.
We then note that for any Pauli operator, $g^2=(g^\prime)^2=1$, and hence $UU^\dagger =1$.  Similar arithmetic gives $UgU^\dagger = g'$.

The relationships with the projectors come from noting that for any state $|\psi\rangle$ in the $S$-codespace, $g |\psi\rangle=|\psi\rangle$.  Thus
\begin{eqnarray}
U|\psi\rangle= \frac{1+g^\prime g}{\sqrt{2}} |\psi\rangle = \frac{1+ g^\prime}{\sqrt{2}}|\psi\rangle=\sqrt{2} P_1 |\psi\rangle\\
U|\psi\rangle= \frac{1+g^\prime g}{\sqrt{2}} |\psi\rangle =  \frac{g-g g^\prime }{\sqrt{2}} |\psi\rangle = \sqrt{2} g P_{-1}  |\psi\rangle.
\end{eqnarray}

To show that $U$ is a Clifford rotation, we first recall that
a Clifford rotation is defined by the property that for any $\sigma$ in the Pauli group on $n$ qubits, $P_n$, $U \sigma U^\dagger =\sigma^\prime \in P_n$.  We take an arbitrary $\sigma\in P_n$, and separately consider the two cases where $\sigma$ commutes or anticommutes with $g g^\prime$. (One of these conditions is always satisfied by any two arbitrary Pauli operators.) If $[\sigma,g g^\prime]=0$, then $U\sigma U^\dagger = \sigma \in P_n$. If $\{\sigma,g g^\prime\}=0$, then
$U\sigma U^\dagger = (\sigma + 2\sigma g g' - \sigma)/2 = \sigma g g'$.  Products of Pauli operators are Pauli operators ($P_n$ is closed under multiplication), so $\sigma g g' \in P_n$.

\section{Fault-tolerant measurement of Pauli operators}\label{ftmeasurement}
A stabilizer measurement requires measuring 
 operators of the form $g=\sigma_1 \sigma_2 \cdots \sigma_m$, which is a product of $m$ Pauli matrices, each acting on a different qubit, labeled by $j=1,\cdots, m$.  
Here we reproduce the argument from Shor \cite{shor} (see also, \cite{Aliferis2006}), showing that this measurement can be done in a fault-tolerant manner.

\begin{figure}[t]
\begin{indented}
\item[]\includegraphics[width =0.8 \columnwidth]{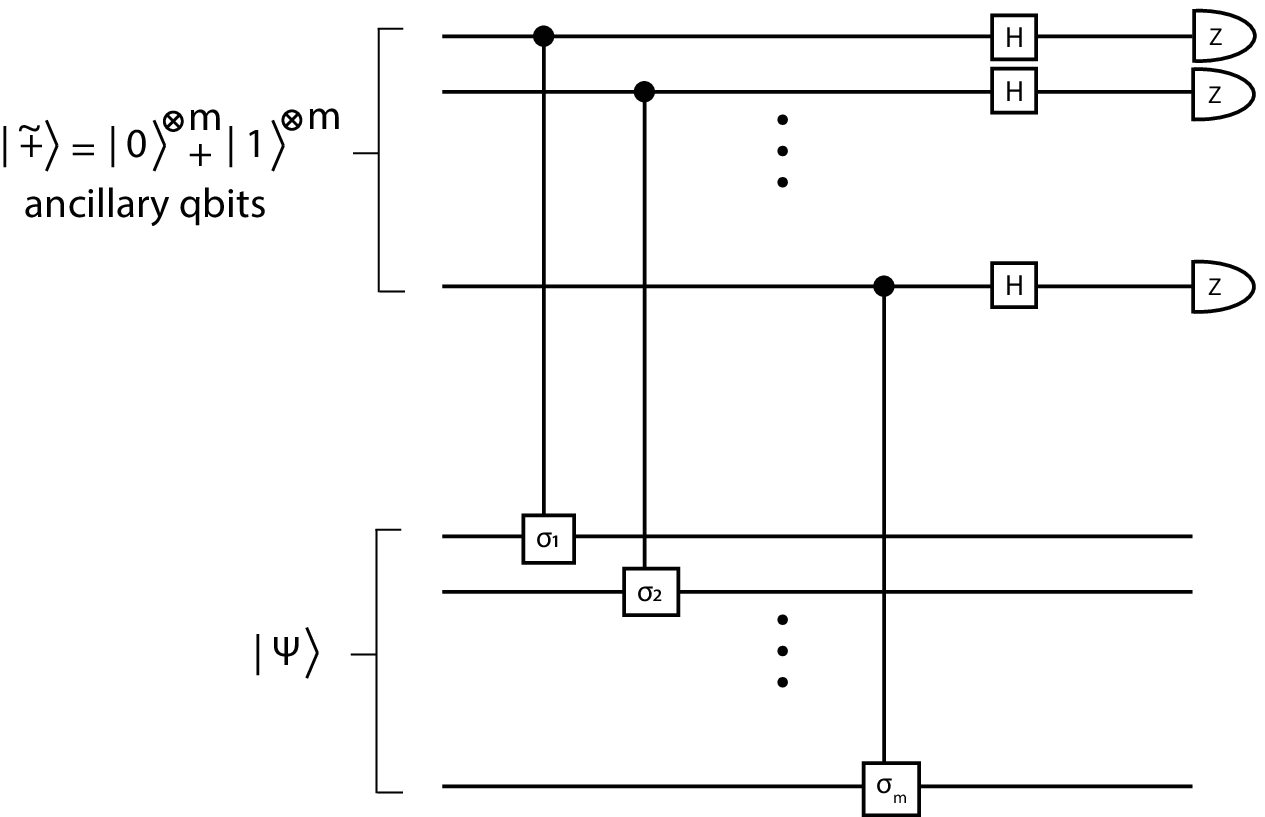}
\end{indented}
\caption{Circuit illustrating the cat-state method \cite{shor,Aliferis2006} for the fault-tolerant projection measurement of a stabilizer generator.  }
\label{fig: ft}
\end{figure}

One first prepares $m$ ancilla qubits in a $m$-qubit cat-state, which is an equal superposition of all ancilla qubits in the $|0\rangle$ state and all ancilla qubits in the $|1\rangle$ state.  As shown in Fig.~\ref{fig: ft}, one then entangles the $j$'th ancilla qubit with the system by applying a control-$\sigma_j$ operator.   A Hadamard gate is applied to each ancilla qubit, then the ancillas are measured in the standard basis.  A measurement having even parity will have projected the encoded state into the $+1$ eigenbasis of $g$.  A measurement having odd parity will have projected the encoded state into the $-1$ eigenbasis of $g$. 
Neglecting normalization, the quantum state of the composite system evolves as 
$
(|0\rangle^{\otimes m}+|1\rangle^{\otimes m})\otimes|\psi\rangle\to 
|0\rangle^{\otimes m}\otimes |\psi\rangle+|1\rangle^{\otimes m}\otimes g|\psi\rangle\to
|e\rangle\otimes (1+g)|\psi\rangle + |o\rangle (1-g) |\psi\rangle,
$
where $|e\rangle$ and $|o\rangle$ are equal weight superpositions of all of the ancilla states with even and odd parity.  Each ancilla qubit interacts with a single, unique physical qubit.  
 
While this scheme prevents errors from propagating, an error on one of the ancilla qubits could be mistaken for an error on one of the system qubits (or could mask such an error).  The standard procedure for addressing this problem is to repeatedly measure the stabilizer.  By comparing subsequent measurements, one can bound the probability of an undetected or misdiagnosed error, yielding a fault-tolerant algorithm.

\section{Trivial Examples}\label{trivial}
To illustrate the mechanics of our algorithm, we carefully apply it to two simple examples.  These are not physically relevant, but they provide a platform for understanding the arithmetic.  

First consider the case where we have $n=2$ physical qubits, storing $k=1$ logical qubits.   Suppose $S$ is generated by $g_0=Z_1$ and $S^\prime$ is generated by $g_0^\prime=Z_2$.  Following our algorithm, we construct the $1\times1$ connectivity matrix $M$, which in this case is equal to zero.  This tells us that our generators are either equal to one-another (which would put them in blocks $G_A$ and $G_A'$), or they are logical operators of the other code (which puts them in blocks $G_B$ and $G_B'$).  Clearly the latter is the case.  The complementary operators are $g_0^{(c)}=X_1$ and $(g_0^\prime)^{(c)}=X_2$.  The sequence of stabilizer codes is then generated by $G_0=\{Z_1\},G_1=\{X_1 X_2\},G_2=\{Z_2\}$.  Each subsequent set of generators differs by exactly one non-commuting element, allowing us to map between them via the the procedure in Sec.~\ref{onestep}.

A somewhat more sophisticated, but equally artificial, example involves $n=3$ physical qubits and $k=1$ logical qubits.  Suppose $S$ is generated by $g_0=Z_1 Z_2, g_1=Z_3$ while $S^\prime$ is generated by $g_0^\prime=Z_1,g_1^\prime=X_2 X_3$,  The connectivity matrix is then
\begin{equation}
M=\left(\begin{array}{cc}
0&1\\
0&1
\end{array}\right).
\end{equation} 
This is made diagonal via the transformation $g_0 \to \bar g_0=g_0 g_1=Z_1 Z_2 Z_3$.  By construction,  $S$ is generated by $\bar g_0$ and $g_1$.  The operators $\bar g_0$ and $g_0^\prime$ are logical operators for the other code (so are in block $G_B$ and $G_B'$), and the complementary operators can be taken to be
$\bar g_0^{(c)}= X_3$ and ${g_0^\prime}^{(c)}=X_1 X_2$.  Following Sec.~\ref{construct}, we then check the anticommutator of these complementary operators with the  elements of $S$ and $S^\prime$.  We find that $\bar g_0^{(c)}$ anticommutes with $g_1$, necessitating the replacement $g_1\to \bar g_1=g_1 \bar g_0$.  
  The sequence of stabilizer generators is then:
$\{Z_1 Z_2 Z_3,Z_1Z_2 \}\to \{X_1 X_2 X_3,Z_1Z_2 \}\to\{Z_1,Z_1Z_2\}\to\{Z_1,X_2X_3\}$.

\section{Code distance during rewiring}\label{fail}
 
The distance of a code $d$ is the minimum number of single qubit errors required to give a non-zero overlap between two basis states in the code-space.  Equivalently, $d$ is the minimum number of single qubit Pauli operators which are needed to construct a logical operator.  In a distance $d$ code one can correct $t=(d-1)/2$ errors.  As described in Sec.~\ref{mindist}, an important question is how the distance of the code evolves along the path in Sec.~\ref{path}.  In that section we established a lower bound on the distance of these codes.  Nonetheless,
in many practical examples (Secs.~\ref{utc}, \ref{top}) the distance of the code is maintained throughout the path.  Here we explicitly construct a mapping between two distance 3 codes for which our algorithm yields an intermediate code of lower distance.

In particular, consider the case where $S_0=S$ is the Steane code defined by the first six stabilizers in the left-hand column of Table \ref{sr}, and $S_2=S^\prime$ is a modified version of the same code where the third and fourth qubit are exchanged -- that is in each stabilizer one makes the substitution  $X_3\to X_4, Z_3\to Z_4, X_4\to X_3, Z_4\to Z_3$.  Both of these are distance 3 codes.  Our algorithm gives an intermediate code $S_1$ which is generated by $\{ Z_1 Z_4 Z_5 Z_7, X_1 X_2 X_5 X_6, X_1 X_3 X_4 X_6, Z_1 Z_3 Z_5 Z_7, Z_1 Z_2 Z_5 Z_6, Z_1 Z_3 Z_4 Z_6\}$.  This code is only distance $1$ as the operator $Z_7$ commutes with all of the generators, but is not itself a stabilizer -- and is therefore a logical operator.  Thus if a ``phase" error occurs on qubit 7 in this intermediate state, the error could neither be detected, nor corrected.

One approach to avoiding this issue is given in \cite{Huang2018}.

\section{Connectivity matrices for mapping between Steane and Reed-Muller Codes}\label{storm}

\begin{table}[t]
\caption{Connectivity matrix for the Steane and Reed-Muller codes.  A one in the row indicates an anticommutation relation, and zeros are inserted where the stabilizer generators are identical.}\label{conn}
\footnotesize

\begin{tabular} {@{}ccccccccccccccc}
\br
& $g_0'$ & $g_1'$ & $g_2'$ & $g_3'$ & $g_4'$ & $g_5'$ & $g_6'$ & $g_7'$ & $g_8'$ & $g_9'$ & $g_{10}'$ & $g_{11}'$ & $g_{12}'$ & $g_{13}'$ \\
\mr 
$g_0$ & & & & & & & & & & 1 & & & &  \\
$g_1$ & & & & & & & & & 1 & & & & &  \\
$g_2$ & & & & & & & & & & & 1 & & &   \\
$g_3$ & & & & & & & & & & & & 0 & &\\
$g_4$ & & & & & & & & & & & & & 0 &\\
$g_5$ & & & & & & & & & & & & & & 0\\
$g_6$ & & & & 1 & & & & & & & & & &   \\
$g_7$  & 1 & & & 1 & & & & & & & & & & \\
$g_8$  & & 1 & & 1 & & & & & & & & & & \\
$g_9$  & 1 & 1 & & 1 & & & & & & & & & &\\
$g_{10}$  & & & 1 & 1 & & & & & & & & & &\\
$g_{11}$  & 1 & & 1 & 1 & & & & & & & & & &\\
$g_{12}$  & & 1 & 1 & 1 & & & & & & & & & &\\
$g_{13}$  & 1 & 1 & 1 & 1 & & & & & & & & & &\\
\br
\end{tabular}

\end{table}

Here we explicitly show how our algorithm is applied to finding a mapping between the Steane and Reed-Muller codes.  As described in  Sec.~\ref{utc}, we first append arbitrary stabilizers to the Steane code so that there are an equal number of stabilizers in both codes (see Table~\ref{sr}).  Then, we form the connectivity matrix and find its diagonal form by mod(2) row reduction.  

The row reduction procedure yields a new set of generators that are used to construct the sequence of measurements described in Sec.~\ref{utc}.   We additionally use column reduction in order to produce the simplest sequence of measurements for mapping in the opposite direction.  We start by creating the connectivity matrix, Table~\ref{conn}, corresponding to the operators in Table~\ref{sr}.  We show ones where the stabilizers anticommute, and zeros are implied at all other locations. We place explicit zeros in the locations where the stabilizers are identical. 

We first note that the first six rows and the last six columns decouple from the others, and we can diagonalize them by simply reordering the rows and columns.  Three of these  generators are identical (and hence belong in block A), $(g_3, g_4, g_5) = (g_{11}', g_{12}', g_{13}')$.  The other elements form anticommuting pairs (and hence belong in block C),   $\{g_0,g_9'\} = \{g_1,g_8' \} = \{g_2,g_{10}' \} = 0$.

We diagonalize the last 8 rows using row reduction mod(2).  First, we rearrange the rows so that the rows with ones in the first column  are at the top, rows with ones in the second column are next, etc.  This yields an ordering $g_{13},g_{11},g_{9},g_{7},g_{12},g_{8},g_{10},g_{6}$.  We add the first row (mod 2) to the second, third, and fourth, to eliminate the ones in the first column, yielding generators $g_{13},g_{11} g_{13},g_{9} g_{13},g_{7} g_{13},g_{12},g_{8},g_{10},g_{6}$.  Table~\ref{rowred} shows the resulting connection matrices after slight reordering.  We then pivot, adding the fifth row $g_{12}$ to the first $g_{13}$, making the first four rows diagonal.  The remaining $1$'s in the last four rows are then readily eliminated by adding rows from this diagonal block (see Table~\ref{rowred}).

\begin{table}
\caption{Connection matrices for intermediate steps of the row reduction used to find the generators needed to map from the Steane code to the Reed-Muller code.\\}\label{rowred}
\begin{indented}
\item[]\begin{tabular}{@{}ll}

\begin{tabular} {@{}ccccc}
\br
& $g_0'$ & $g_1'$ & $g_2'$ & $g_3'$ \\
\mr 
$g_{13}$ & 1 & 1 & 1 & 1 \\
$g_{11}g_{13}$ &  & 1 &  &  \\
$g_{9}g_{13}$ &  &  & 1 &  \\
$g_{6}$ & &  &  & 1 \\
$g_{12}$ & & 1 & 1 & 1 \\
$g_{8}$ & & 1 &  & 1 \\
$g_{7}g_{13}$ & & 1 & 1 & \\
$g_{10}$ & & & 1 & 1 \\
\br
\end{tabular}
&
\begin{tabular} {@{}ccccc}
\br
& $g_0'$ & $g_1'$ & $g_2'$ & $g_3'$ \\
\mr 
$g_{13}g_{12}$ & 1 & &  &  \\
$g_{11}g_{13}$ &  & 1 &  &  \\
$g_{9}g_{13}$ &  &  & 1 &  \\
$g_{6}$ & &  &  & 1 \\
$g_{12} g_{6} g_{9} g_{11}$ & &  &  &  \\
$g_{8} g_6 g_{11} g_{13}$ & &  &  &  \\
$g_{7}g_{13} g_{11} g_9$ & &  & & \\
$g_{10} g_6 g_9 g_{13}$ & & &  &  \\
\br
\end{tabular}
\end{tabular}
\end{indented}
\end{table}

Thus we have four new elements for block C: $\{g_{13}g_{12},g_0^\prime\}=\{g_{11}g_{13},g_1^\prime\}=\{g_9g_{13},g_2^\prime\}=\{g_6,g_3^\prime\} = 0$.  The remaining generators either belong to block A (meaning they are equal to stabilizers of the other code) or block B (in which case they are logical operators for the other code).  Given that the two codes share a logical operator $Z=\prod_j Z_j$, block B must be empty, and the rest of the generators are in block A.

To find the most convenient sequence to convert backwards from the Reed-Muller code to the Steane code, we follow the same procedure but use column reduction.  We reorder the rows as shown in Table~\ref{colred}, and add columns together (mod 2) in order to reduce the lower four rows to diagonal form.  To simplify the expressions we also reorder the rows.  Finally, we use row reduction to eliminate all ones in the upper four rows.  This establishes a simple mapping backwards.  To map a state from the codespace of the Reed-Muller code to that of the Steane code, we measure: $g_0$, $g_1$, $g_2$, $g_6$, $g_8$, $g_9$, $g_{10}$.

\begin{table}[h]
\caption{\label{colred}Connection matrices for intermediate steps of the column reduction used to find the generators needed to map from the  Reed-Muller code to the Steane code.\\}

\begin{indented}
\item[]\begin{tabular}{@{}lll}
\begin{tabular} {@{}lcccc}
\br
& $g_0'$ & $g_1'$ & $g_2'$ & $g_3'$ \\
\mr 
$g_{13}$ & 1 & 1 & 1 & 1 \\
$g_{11}$ & 1 & & 1 & 1 \\
$g_{9}$ & 1 & 1 &  & 1 \\
$g_{7}$ & 1 &  &  & 1 \\
$g_{12}$ & & 1 & 1 & 1 \\
$g_{8}$ & & 1 &  & 1 \\
$g_{10}$ & & & 1 & 1 \\
$g_{6}$ & & & & 1 \\
\br
\end{tabular}
&
\begin{tabular} {lcccc}
\br
& $g_0'$ & $g_1' g_0^\prime$ & $g_2'$ & $g_3' g_2^\prime g_1^\prime$ \\
\mr 
$g_{13}$ & 1 &  & 1 & 1 \\
$g_{11}$ & 1 & 1& 1 &  \\
$g_{12}$ & & 1 & 1 &1  \\
$g_{7}$ & 1 & 1 &  & 1 \\
$g_{9}$ & 1 &  &  &  \\
$g_{8}$ & & 1 &  &  \\
$g_{10}$ & & & 1 &  \\
$g_{6}$ & & & & 1 \\
\br
\end{tabular}
\end{tabular}\\[0.1in]
\begin{tabular} {@{}lcccc}
\br
& $g_0'$ & $g_1' g_0^\prime$ & $g_2'$ & $g_3' g_2^\prime g_1^\prime$ \\
\mr 
$g_{13} g_9 g_{10} g_{6}$ &  &  &  &  \\
$g_{11} g_9 g_8 g_{10}$ &  & &  &  \\
$g_{12} g_8 g_{10} g_6$ & &  &  &  \\
$g_{7} g_9 g_8 g_6$ &  &  &  &  \\
$g_{9}$ & 1 &  &  &  \\
$g_{8}$ & & 1 &  &  \\
$g_{10}$ & & & 1 &  \\
$g_{6}$ & & & & 1 \\
\br
\end{tabular}
\end{indented}

\end{table}

\end{document}